# Superconductivity in compressed SnPS$_3$

Binbin Yue[1]*, Wei Zhong[1], Ting Wen[1], Yonggang Wang[1], Hui Yu[2,3], Xiaohui Yu[2,3,4], Fang Hong[2,3,4]*

[1] *Center for High Pressure Science & Technology Advanced Research, 10 East Xibeiwang Road, Haidian, Beijing 100094, China*

[2] *Beijing National Laboratory for Condensed Matter Physics, Institute of Physics, Chinese Academy of Sciences, Beijing 100190, China*

[3] *School of Physical Sciences, University of Chinese Academy of Sciences, Beijing 100190, China*

[4] *Songshan Lake Materials Laboratory, Dongguan, Guangdong 523808, China*

*Email: yuebb@hpstar.ac.cn; hongfang@iphy.ac.cn

**Abstract**

Metal phosphorous trichalcogenides, $MPX_3$, is a group of van der Waals materials with rich electronic properties and even exotic magnetic behavior. These properties can be well manipulated by pressure/strain via effective control of interlayer interaction, lattice parameters and crystal structure. Superconducting transition has been observed in compressed $FePSe_3$. However, it is the only one superconductor reported in the large $MPX_3$ family. Is it possible to achieve superconducting transition in other $MPX_3$ compounds, especially in a trisulfide compound? In this work, we tentatively compressed the $SnPS_3$ (an insulator with large band gap at ambient condition) up to 48.9 GPa, and managed to achieve the superconducting transition above 31.7 GPa with $T_c$ ranging from ~2.2 K to ~2.8 K. The upper critical field is estimated to be ~3.03 T at 40.5 GPa. Optical absorption measurements together with Raman spectroscopy show a series of transitions under pressure, which is well consistent with the electric transport results. This work provides direct experimental evidence that $SnPS_3$ undergoes an insulator-metal transition near 31.7 GPa. More importantly, it demonstrates that superconductivity can exist in $MPS_3$ compounds, which not only further enriches the electronic properties of this kind of material but also paves a new avenue to explore the abundant emergence phenomena in the whole $MPX_3$ family, and it also benefits the study of superconductor mechanism.

**Introduction**

The group of metal phosphorous trichalcogenides, $MPX_3$, is a large family of layered materials with metal cations stabilized $[P_2X_6]^{4-}$ framework layers weakly bonding each other via van der Waals interactions [1-4]. The metal M includes transition metals (*e.g.*, V, Cr, Mn, Fe, Co, Ni, Cu, Zn, Cd, Hg, Pd, and Ag) as well as post-transition metals (*e.g.*, In, Ga, Sn and Pb) and alkali metals (*e.g.*, Mg and Ca). For the chalcogen atoms X, only S and Se are currently reported. Most of these compounds adopt a monoclinic or rhombohedral lattice structure derived from $CdCl_2$ and $CdI_2$ structural type with different stacking of individual layers. All materials in this group are semiconductors with band gap from 1.3 to 3.5 eV [5-7], allowing optoelectronic applications in a broad wavelength range. In addition, the magnetic properties and uncommon intercalation-substitution behavior in various $MPX_3$ compounds have also attracted numerous attentions with potential application in spintronic devices [4,8], batteries [9] and catalysis [10].

Previous works demonstrated that the electronic properties of $MPX_3$ can be tuned via chemical doping [11,12], temperature [13], and pressure/strain [14-21]. The introduction of copper in $SnPS_3$ has a substantial impact on the shape smearing of the ferroelectric domains as well as on the behavior of the dielectric permittivity [12]. Trisulfide compounds with M = Mn, Fe, Co, and Ni are paramagnetic at high temperatures and undergo a transition to antiferromagnetic state upon cooling [13]. Under applied pressure, the crystal structure along with related band structure of a crystalline material can be finely tuned in a clean and controllable manner without introducing chemical disorder. High-pressure studies on several $MPX_3$ materials, including $FePS_3$, $FePSe_3$, $MnPS_3$, $MnPSe_3$, $NiPS_3$, $V_{0.9}PS_3$ have shown Mott metal-insulator transition (MIT) [14-21]. Apart from the pressure induced MIT effect, spin-crossover and a large volume collapse has also been observed in iron and manganese-based $MPX_3$ compounds [14-19]. More interestingly, superconductivity (SC) emerges at around 3 ~ 5 K in the nonmagnetic high-pressure phase of $FePSe_3$ above ~9 GPa [15]. Pressure induced SC transition has been reported in many metal chalcogenides, like $MoS_2$ [22], $SnSe_2$ [23], and $HfS_3$ [24], et. al. However, $FePSe_3$ is the only $MPX_3$ compound showing SC transition. It is interesting to explore the possibility of SC transition in other $MPX_3$ compounds, especially in a $MPS_3$ compound.

Here, we reported the pressure induced SC transition in $SnPS_3$. This compound is a very good ferroelectric material with a band gap of ~2.3 eV [25,26] and crystallizes in a monoclinic *Pn* structure at ambient condition [27,28]. Under low pressure, it has a second order transition from the ferroelectric phase to a paraelectric phase with a *P2$_1$/n* lattice above ~ 0.2-0.5 GPa [29,30]. Previous room temperature electric resistivity measurements show a gradual and reversible ten orders of magnitude decreasing by pressurizing $SnPS_3$ up to 20 GPa [31], suggesting possible metallization behavior. However, there is no direct temperature dependent electric transport measurement to confirm such an electronic transition. By combining electric transport measurements, Raman spectroscopy, UV-VIS-NIR and infrared absorption spectroscopy, we successfully revealed the pressure effect on the electronic and structure properties of $SnPS_3$. The low temperature transport measurement shows that $SnPS_3$ is always a semiconductor up to 31.7 GPa, near which it starts to become a semimetal or bad metal.

Such a metallization process was also confirmed by the disappearance of Raman and infrared vibration modes. Above 31.7 GPa, $SnPS_3$ started to undergo a superconducting transition near 2.2 K, which is slightly enhanced to ~2.8 K at 48.9 GPa, the highest pressure in this work. The superconductivity is further confirmed by the magnetic field effect, and the upper critical field is estimated to be 3.03 T by fitting the $T_c$-$H$ curve via the Ginzburg-Landau (G-L) equation. The optical absorption measurement tracked the trend of pressure dependent band gap, and provided a relatively systematic information on the electronic behavior of the compressed $SnPS_3$. The observation of superconductivity in $SnPS_3$ will arouse the passion to explore the pressure induced superconductivity in the large $MPX_3$ family.

**Experiments**

The standard four-probe electrical resistance measurement was carried out under high pressure up to 48.9 GPa, and the experiment was conducted in a commercial cryostat from 1.7 K to 300 K by a Keithley 6221 current source and a 2182A nanovoltmeter. Two opposing anvils with 300 µm culets were placed in a BeCu alloy diamond anvil cell (DAC) to generate high pressure. A thin single crystal sample was loaded into the sample chamber in a rhenium gasket with c-BN insulating layer, and a ruby ball was loaded to serve as an internal pressure standard.

The high-pressure Raman spectra were collected using a Renishaw Micro-Raman spectroscopy system equipped with a second-harmonic Nd:YAG laser (operating at 532 nm). The laser power was maintained at relatively low power level to avoid overheating the sample during measurement. In-situ high-pressure UV-VIS-NIR absorption spectroscopy was performed on a home-designed spectroscopy system (Ideaoptics, Shanghai, China) with a 405 nm laser. The high-pressure infrared experiments were performed at room temperature on a Bruker VERTEX 70v infrared spectroscopy system with HYPERION 2000 microscope. The spectra were collected in absorption mode in the range of 500-8000 cm$^{-1}$ with a resolution of 4 cm$^{-1}$ through a ~50×50 μm$^2$ aperture. In all spectroscopy measurements, type IIa diamond anvils with low fluorescent signal were used and KBr was used as pressure medium. Pressure was determined from the shift of R1-R2 ruby fluorescence lines [32].

**Results and discussions**

Electric transport experiments on SnPS$_3$ were carried out up to 48.9 GPa. Resistance-temperature (*R-T*) results are presented in Fig. 1, and several transitions can be observed in this pressure range. The initial resistance of the sample is too large to be detected by our multimeter, indicating its insulating nature. When pressure is increased to 6.3 GPa, room-temperature resistance becomes measurable with a value of ~10.6 KΩ. This suggests that there might be a structural transition near this pressure point, which greatly improves the conductivity of the sample. However, the resistance increases rapidly upon cooling and becomes out of range when the temperature is lower than ~190 K. The resistance is strongly reduced with applied pressure up to 8.5 GPa. Above this pressure point, the sample resistance around 90 K drastically drops more than four orders of magnitude at 12.2 GPa, and resistance values in the whole temperature range were collected, which might be the hint of another structural transition. Above 15.7 GPa, the decreasing trend of the resistance in the whole temperature range slows down but the *R-T* curves keeps its typical semiconducting nature up to 29.2 GPa. From 31.7 GPa, a sudden drop can be observed around 2.2 K, indicating a possible superconducting (SC) transition. This SC transition is enhanced with higher pressure, the resistance reaches to zero above 38.7 GPa, and $T_c$ slightly increases to ~ 2.8 K at 48.9 GPa. The inset in Fig.1(d) shows the photo of sample inside DAC at 48.9 GPa.

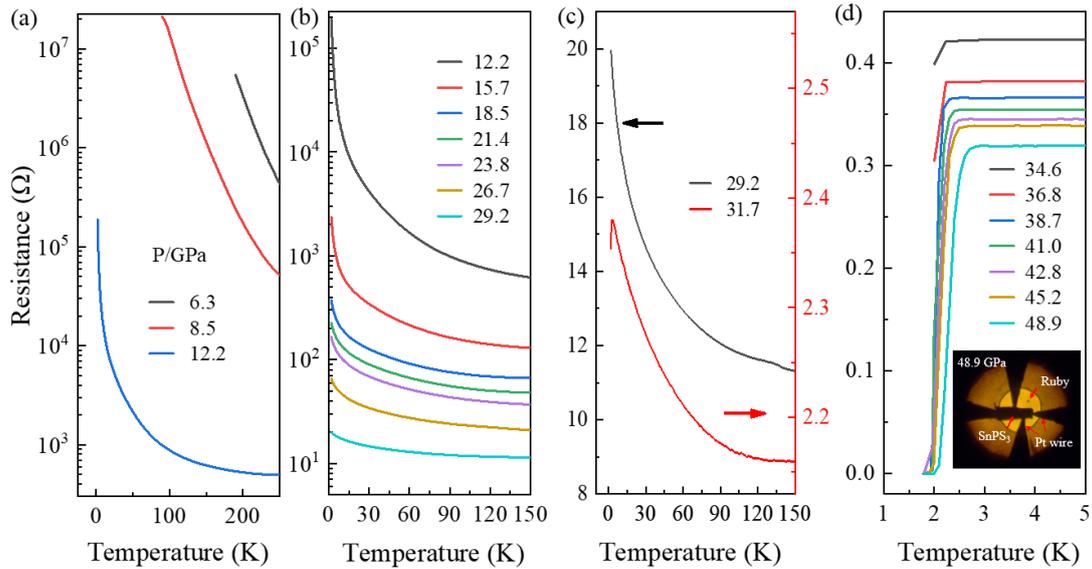

**Figure 1** The electric transport properties of SnPS$_3$ under pressure up to 48.9 GPa. *R-T* curves in (a) 6.3~12.2 GPa, (b) 12.2~29.2 GPa show semiconducting behavior. (c) Sharp drop appears at ~ 2.2 K from 31.7 GPa. (d) SC transition temperature increases with pressure up to 48.9 GPa. Inset in (d) is a photo showing the standard four-probe electrical resistance measurement set up in the DAC chamber.

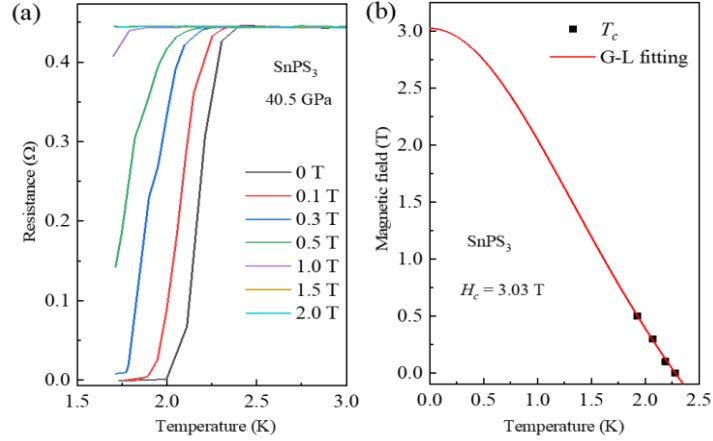

**Figure 2** The magnetic effect on the superconductivity in SnPS$_3$ at 40.5 GPa. (a) R-T curves at various magnetic fields near the superconducting transition region, (b) upper critical magnetic field - temperature phase diagram. The solid line in (b) represents G-L equation fitting.

To further confirm the pressure-induced superconductivity in SnPS$_3$, we measured the temperature dependent resistance at 40.5 GPa under external magnetic field up to 2 T. as shown in Fig. 2 (a). It can be seen clearly that $T_c$ is suppressed to lower temperature with increasing magnetic field, which is typical for a bulk superconducting transition. The SC transition completely disappears under 1.5 T within our lowest temperature limit 1.7 K. The upper critical field $\mu_0H_{c2}$ as a function of critical temperature $Tc$ is plotted in Fig. 2 (b). By fitting the $\mu_0H_{c2}(T)$ with the G-L equation [33], the zero-temperature $\mu_0H_{c2}(0)$ is estimated to be 3.03 T, which is lower than the Bardeen-Copper-Schrieffer (BCS) weak-coupling Pauli paramagnetic limit of $\mu_0H_p = 1.84T_c = 4.2$ T for $T_c = 2.28$ K, suggesting the absence of Pauli pair breaking [34].

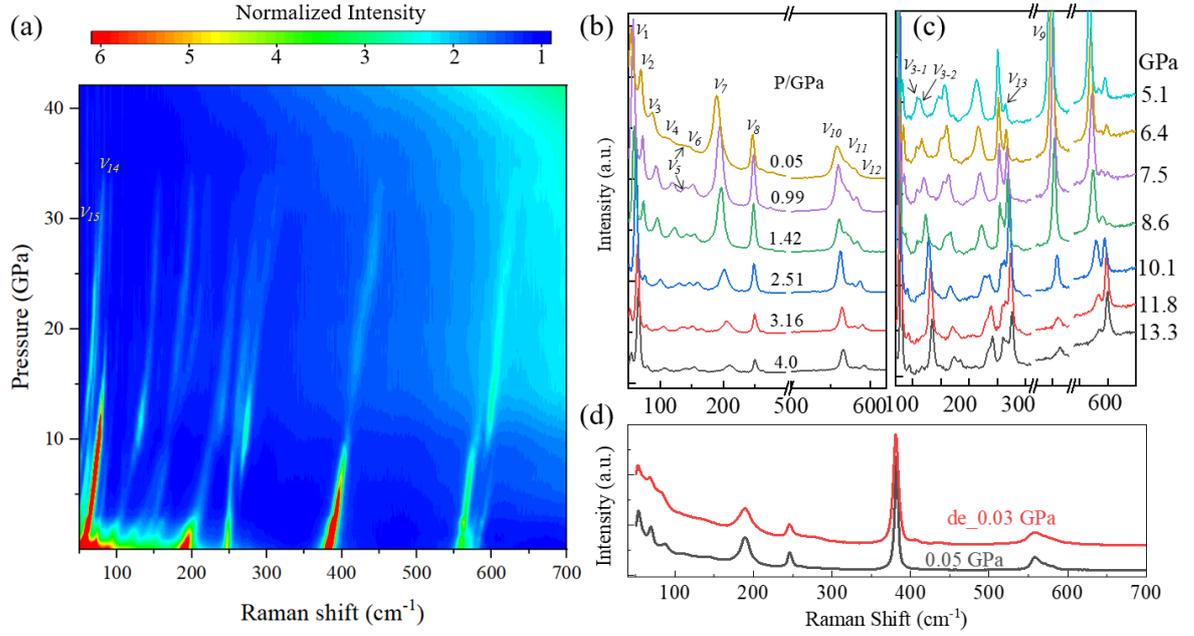

**Figure 3** Raman spectra of SnPS$_3$ under high pressure. (a) Spectra 2D plot shows a series of phase transitions during compression up to 42.1 GPa. 1D spectra in the pressure range of 0-4 GPa (b) and 5.1-13.3 GPa (c). (d) Comparation of spectra measured before and after pressure treatment.

Raman spectroscopy covering frequencies from 50 to 700 cm$^{-1}$ were used to reveal the structural transitions of SnPS$_3$ under compression (Fig. 3). Spectra 2D plot in Fig. 3(a) shows clearly a series of transitions up to 42.1 GPa. Around 12 modes can be observed at 0.05 GPa in the whole frequency range, as labeled in Fig. 3(b) and 3(c). Due to the strong influence of the intermolecular interactions, straightforward eigenvector assignment by frequency comparisons is not allowed for Raman modes below 240 cm$^{-1}$. According to the theoretical calculation results of Smirnov *et al.* [35], modes below 90 cm$^{-1}$ involve primarily the displacement of Sn ions and (P$_2$S$_6$)$^{4-}$ groups, and three modes (54, 69 and 87 cm$^{-1}$) have been observed in this range. With applied pressure, all peaks shift to right and two more peaks ($V_{14}$ and $V_{15}$) can be detected in this range.

Raman modes lying between 90 and 140 cm$^{-1}$ represent primarily $(P_2S_6)^{4-}$ rotations [35]. Peaks in this range are very weak at 0.05 GPa and become much stronger from 0.99 GPa, which might be addressed to the known second order phase transition in SnPS$_3$ as mentioned above. Raman modes lying between 140 and 240 cm$^{-1}$ are essentially $(P_2S_6)^{4-}$ rocking and bending modes [35] and one strong peak could be found at 189 cm$^{-1}$. Modes above 240 cm$^{-1}$ correspond to the high-frequency internal modes of the $(P_2S_6)^{4-}$ with a strongest peak at 382 cm$^{-1}$ representing the S$_3$P-PS$_3$ vibration [36]. The region between 550 and 600 cm$^{-1}$ contains the P-S valence vibration and also shows slight change during the first phase transition.

With increasing pressure, several transitions can be observed. From 2.51 GPa, $V_2$ and $V_{11}$ modes decay a lot. From 5.1 GPa, a small peak at 263 cm$^{-1}$ ($V_{13}$) appears, combining with the enhancement of mode $V_1$, the splitting of mode $V_3$ to $V_{3-1}$ and $V_{3-2}$, and the disappearance of mode $V_6$. This structure transition corresponds to the resistance decrease of the sample near this pressure range. Above 8.55 GPa, the mode $V_9$ is weakened a lot while modes $V_{3-2}$, $V_{13}$ and $V_{11}$ are all enhanced. In the low-frequency range, the disappearance of mode $V_2$ can also be observed near this pressure point. This phase transition was not reported before, but is quite clear here and can be correlated with the conductivity increase in the low-temperature range near this pressure. From 14.6 GPa, the mode $V_1$ becomes weak and disappears at 20 GPa. From 25.2 GPa, mode $V_{14}$ starts to decay and almost disappears at 30.2 GPa. Finally, all peaks disappear above 35.1 GPa, indicating possible metallization of the sample. This is consistent with previous results reported by Ovsyannikov *et al.* [37]. Raman results also demonstrate that all electronic transitions are corresponding to relevant (micro)structural transitions.

The electronic structure has been detected by the UV-VIS-NIR optical absorption spectroscopy at ambient condition and under high pressure (Fig. 4). At ambient condition, $SnPS_3$ shows a clear absorption edge at ~517 nm, with an indirect and direct band gap fitting to be 2.23 and 2.37 eV, respectively. This is consistent with previous theoretical and experimental results [25,26]. Under applied pressure of ~0.4 GPa, the absorption edge shows obvious red shift to ~537 nm, with a direct band gap of 2.24 eV. This large drop of band gap is corresponding to the structural phase transition observed in Raman data. With further compression, the absorption edge shifts to higher wavelength smoothly until 10.8 GPa. Then it moves much faster and enters near infrared range at 16.6 GPa. At 20.2 GPa, the band gap decreases to 0.54 eV and then moves out of the NIR detecting range under higher pressure. During compression, the color of the sample also changes from yellow to red and then black from 1.4 to 6.6 GPa, which shows direct evidence of the rapid shrinkage of the energy band gap. During pressure release, the sample color changes from back to yellow, and the absorption edge also move back to 544 nm at 0.5 GPa, similar with that at 0.4 GPa during compression (Fig. 4a), agreeing well with the reversible nature of the phase transition.

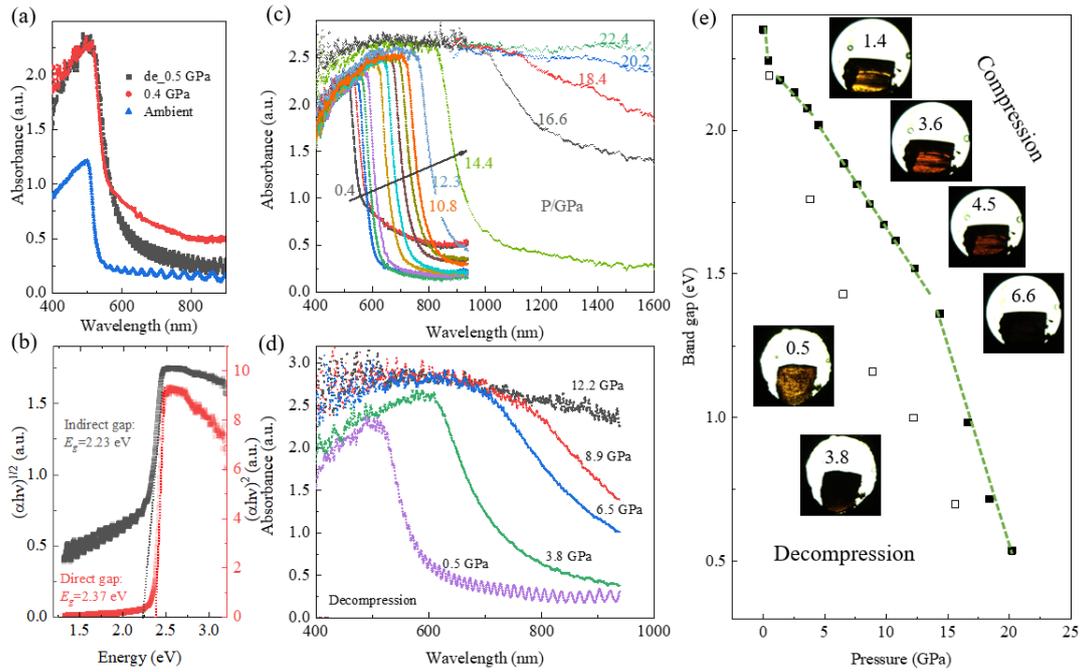

**Figure 4** Optical and electronic properties of SnPS$_3$ under pressure. (a) The optical absorption spectroscopy of the sample before and after pressure treatment. (b) Direct and indirect band gap determination at ambient condition. The UV-VIS-NIR absorption spectra during compression up to 22.4 GPa (c) and decompression (d). (e) The pressure dependent optical band gap. Insets are photographs of the SnPS$_3$ sample at selected pressures inside diamond anvil cell.

Mid-infrared absorption spectroscopy measurements were carried out to get the electronic information under further compression and results are displayed in Fig. 5. The absorbance spectrum at 0.4 GPa contains clear fringes caused by the Fabry-Pérot effects in the optically flat sample and in the KBr and didn't change too much up to 11.2 GPa. Above this pressure point, the absorbance shows slight increase with pressure until 20.2 GPa. Then rapid jump of absorbance and the decay of fringes can be observed from 22.5 GPa, indicating an electronic structure change here. Above 30 GPa, fringes can no longer be observed which indicates the metallization of the sample. This agrees pretty well with the Raman results.

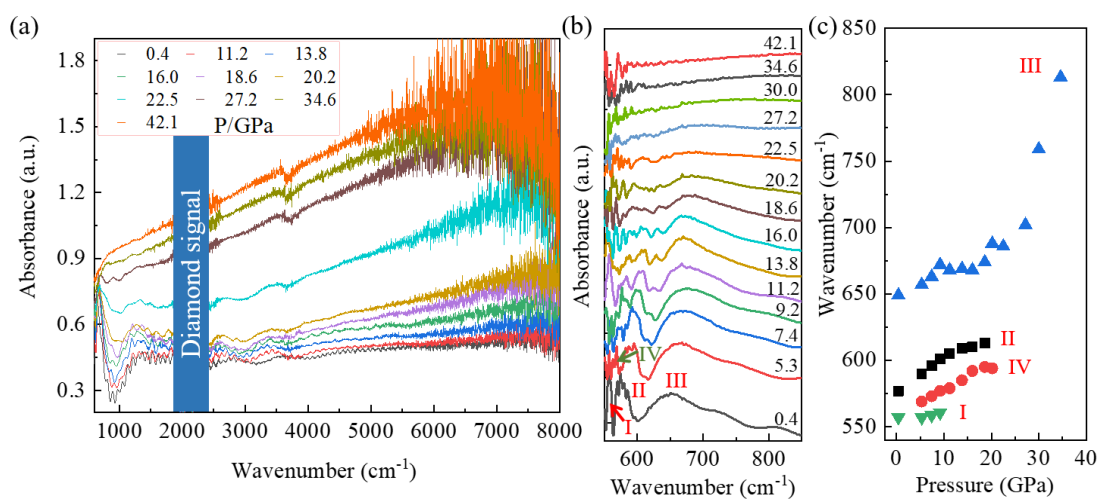

**Figure 5**. Mid-infrared absorption spectra of SnPS$_3$ under pressure up to 42.1 GPa. (a) Selected spectra show the absorbance increase trend with pressure. (b) Enlarge of the low wavenumber range showing several infrared modes. (c) Pressure dependences of frequencies of infrared modes during compression.

If we take a closer look at the low frequency region, several absorption peaks can be observed. Below 600 cm$^{-1}$, it exhibits one sharp peak at 557 cm$^{-1}$ and another with its maximum at ~577 cm$^{-1}$. This is consistent with the IR spectrum reported previously [38,39] and can be assigned to the (P-S) stretching vibration in the "PS$_3$" unit. Besides, a strong broad asymmetrical line with its maximum at 649 cm$^{-1}$ could be seen beyond 600 cm$^{-1}$. With the increase of pressure, a new mode appears at ~ 569 cm$^{-1}$ from 5.3 GPa and the it shows blue shift together with the other modes. The mode at 557 cm$^{-1}$ shows a much smaller pressure coefficient and disappear from 11.2 GPa, together with the softening of the mode at 649 cm$^{-1}$, which further proves the possible phase transition near this pressure point. Transition at 20.2 GPa is proved by the disappearance of mode II and mode IV. From 22.5 GPa, mode III moves much faster and finally disappears from 34.6 GPa when the sample enters a metallic state.

Based on spectroscopy and electric transport measurements, the pressure-temperature (*P-T*) phase diagram of SnPS$_3$ has been plotted and is shown in Fig.6. At low pressure, SnPS$_3$ is an insulator with an unmeasurable resistance and a band gap larger than 2 eV. From ~ 6.3 GPa, it enters a new semiconducting phase with reduced band gap. Due to the continuous structural transitions under higher pressure, resistance of the sample drops with different rate in different pressure range while still shows a semiconductor nature. From 31.7 GPa, superconductivity started to appear near 2.2 K upon cooling, together with the sudden drop of $R_{300\,K}$. Within the highest-pressure limit in this work, $T_c$ is quite robust and slightly increases to ~ 2.8 K at 48.9 GPa.

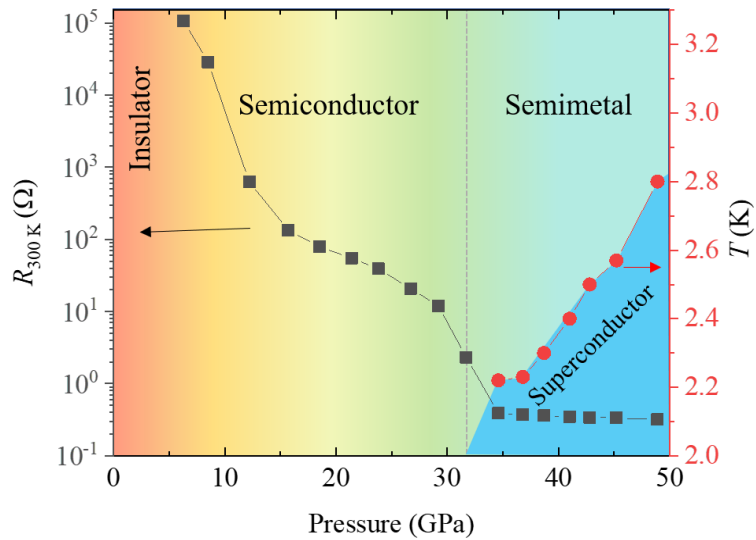

**Figure 6**. Pressure-temperature phase diagram of SnPS$_3$. Black squares represent the resistance of SnPS$_3$ at 300 K. Red circles show the $T_c$. Dash line is plotted for eye guide, showing the approximate boundary between semiconductor and semimetal states.

Motivated by current work, it is promising to explore the superconductivity in other compounds since there are dozens of MPX$_3$ compounds in this family. Meanwhile, in many magnetic MPX$_3$ compounds, the correlation between magnetism and possible superconductivity is still unknown, which is also an important research topic. By changing the M or X element, the chemical pressure effect and electron/hole doping effect will be introduced beyond the (quasi)hydrostatic pressure, which is another manipulation tool to study the superconductivity mechanism. In addition, the phase transition routes of different MPX$_3$ systems may be different from each other, and how the structure affects the superconductivity is another important research topic as well. Therefore, MPX$_3$ provides an ideal platform to study the emergent physical properties and the underlying mechanism by incorporating the high pressure, and related research is still at the early stage.

**Conclusion**

In summary, the pressure-induced superconductivity in SnPS$_3$ was investigated by combining the electric transport and spectroscopy measurements. After a series of electronic and structural transitions, superconductivity was observed above 31.7 GPa. The superconducting behavior is quite robust under pressure and the Tc slightly increases from ~2.2 K to ~2.8 K up to 48.9 GPa. The critical field is estimated to be ~3.03 T. SnPS$_3$ should be the first sulfide compound with superconductivity confirmed by experiment in this family. This work not only provided detailed study on the electronic and structural phase transitions in SnPS$_3$ under high pressure, but also stimulate the effort to explore pressure induced superconductivity in the large MPX$_3$ family.

**Acknowledgement**

This work was supported by the National Natural Science Foundation of China (Grant No. 12004014, U1930401) and Major Program of National Natural Science Foundation of China (22090041).

**Author contribution**

arXiv :2111.02060 (2021).

[25] R. V. Gamernyk, Y. P. Gnatenko, P. M. Bukivskij, P. A. Skubenko, and V. Y. Slivka, J. Phys.: Condens. Matter **18**, 5323 (2006).

[26] K. Kuepper, B. Schneider, V. Caciuc, M. Neumann, A. V. Postnikov, A. Ruediger, A. A. Grabar, and Y. M. Vysochanskii, Phys. Rev. B **67**, 115101 (2003).

[27] G. Dittmar and H. Schäfer, Z. Naturforsch. B **29**, 312 (1974).

[28] B. Scott, M. Pressprich, R. D. Willet, and D. A. Cleary, J. Solid State Chem. **96**, 294 (1992).

[29] O. Andersson, O. Chobal, I. Rizak, V. Rizak, and V. Sabadosh, Phys. Rev. B **83**, 134121 (2011).

[30] Y. Tyagur and I. Tyagur, High Pressure Res. **28**, 607 (2008).

[31] V. V. Shchennikov, N. V. Morozova, I. Tyagur, Y. Tyagur, and S. V. Ovsyannikov, Appl. Phys. Lett. **99**, 212104 (2011).

[32] H. K. Mao, J. Xu, and P. M. Bell, J. Geophys. Res. Solid Earth **91**, 4673 (1986).

[33] J. A. Woollam, R. B. Somoano, and P. O'Connor, Phys. Rev. Lett. **32**, 712 (1974).

[34] A. M. Clogston, Phys. Rev. Lett. **9**, 266 (1962).

[35] M. B. Smirnov, J. Hlinka, and A. V. Solov'ev, Phys. Rev. B **61**, 15051 (2000).

[36] P. H. M. van Loosdrecht, M. M. Maior, S. B. Molnar, Y. M. Vysochanskii, P. J. M. van Bentum, and H. van Kempen, Phys. Rev. B **48**, 6014 (1993).

[37] S. V. Ovsyannikov, H. Gou, N. V. Morozova, I. Tyagur, Y. Tyagur, and V. V. Shchennikov, J. Appl. Phys. **113**, 013511 (2013).

[38] G. Kliche, J. Solid State Chem. **51**, 118 (1984).